# Microwave Loss Reduction in Cryogenically Cooled Conductors

R. Finger · A. R. Kerr

**Abstract** Measurements of microwave attenuation at room temperature and 4.2 K have been performed on some conductors commonly used in receiver input circuits. The reduction in loss on cooling is substantial, particularly for copper and plated gold, both of which showed a factor of 3 loss reduction. Copper passivated with benzotriazole shows the same loss as without passivation. The residual resistivity ratio between room temperature and 4.2 K, deduced from the measurements using the classical skin effect formula, was smaller than the measured DC value to a degree consistent with conduction in the extreme anomalous skin effect regime at cryogenic temperatures. The measurements were made in the 5–10 GHz range. The materials tested were: aluminum alloys 1100-T6 and 6061-O, C101 copper, benzotriazole treated C101 copper, and brass plated with electroformed copper, Pur-A-Gold 125-Au soft gold, and BDT200 bright gold.

## 1 Introduction

Input circuit loss is an important factor limiting the sensitivity of microwave and millimeter-wave receivers. Recently, there has been interest in the possibility of reducing the loss of cryogenic microwave components by using conductors whose resistivity is known to decrease substantially on cooling, and even high temperature superconductors.

High temperature superconductors are used in some very low loss microwave circuits, but it is not clear whether, in a typical receiver application, they would have a real

R. Finger (✉) · A. R. Kerr
National Radio Astronomy Observatory, 1180 Boxwood Estate Rd, Charlottesville, VA 22903, USA
e-mail: rfinger@nrao.edu

advantage over pure metals whose DC resistivity at low temperatures can be more than an order of magnitude smaller than at room temperature.

For high-Q circuits, such as narrow-band filters, superconductors have an advantage over normal metals, but in the usual cryogenic input circuitry of a radio astronomy receiver—mode transducers, polarizers, transmission lines, hybrids, couplers, all of which have low Q—the low resistivity of pure normal metals may result in loss of an acceptably low level. The practical advantages of using plated or machined metal conductors, rather than high temperature superconducting ceramic materials which must be deposited on high dielectric constant substrates, are apparent.

While aluminum, copper and gold-plated conductors are commonly used in the input circuits of receivers, the improvement in performance due to the reduction in loss at cryogenic temperatures is not well documented. DC measurements [1] indicate that the change of resistivity of these metals on cooling is strongly dependent on material purity and the internal stress resulting from machining and thermal treatment.

This paper reports measurements in the 5–10 GHz range of the loss of some machined and plated conductors at room temperature and in liquid helium (4.2 K). As corrosion is an important consideration for bare cooper, we also measured the loss of copper conductors passivated using benzotriazole [2], a simple dip process developed by the U. S. Navy which has been found not to increase the loss significantly in WR-10 copper waveguides at ~100 GHz [3].

## 2 Measurement setup

Capacitively coupled coaxial resonators were made from seven materials commonly used in microwave devices: aluminum alloys 1100-T6 and 6061-O, C101 copper, benzotriazole treated C101 copper, and brass plated with electroformed copper, Pur-A-Gold 125-Au soft gold, and BDT200 bright gold [4]. The plating thickness was ~7.3 μ$m$ for gold and ~6.3 μ$m$ for copper, which is approximately 7 skin-depths at 5.5 GHz and 9 skin-depths at 7.6 GHz at room temperature. The resonator body and center conductor were made of the same material in each case.

As shown in Fig. 1, a transmission line resonator, supported on high density polyethylene (HDPE) pins, is capacitively coupled to a through-line to give a series of resonant transmission minima in the 5–10 GHz window. The through-line and resonator both have a characteristic impedance of 50 ohms. Stainless steel input and output cables are coupled to the circuit via SMA connectors. A heat strap (not shown in the figure) protruding from the end of the resonator assembly dips into the liquid helium and allows the circuit to be cooled to 4.2 K without the cavity filling with liquid.

Warm and cold insertion loss data were measured at three resonant frequencies using an HP8722D vector network analyzer. Because the resonances were very sharp, they were measured in separate narrow frequency windows, 5.5–5.9 GHz, 7.3–7.9 GHz and 9.1–9.9 GHz. Taking 1601 frequency points in each window allowed the off-resonance baseline and the resonance characteristics to be clearly determined.

Using the microwave circuit simulator *Microwave Office,* a transmission line model of the resonator, Fig. 2, was optimized to fit the measured data. The capacitors connected to ground in the figure account for the capacitance added by the HDPE spacers and the fringing capacitance at the end of the resonator. The resistor accounts for the small conductivity of the HDPE. The four free parameters are: the resonator length (*length*), the coupling capacitance (*C* of C1), the support piece capacitance (*HDPE_C*), and the resonator loss (Res_loss), as defined in the figure. The Simplex algorithm was used to find the parameter values that minimize the cost function $C = \Sigma |x_i - m_i|^2$, $x_i$ being the measured insertion loss at the $i$-th frequency and $m_i$ the corresponding insertion loss of the model.

The optimization was done simultaneously over all frequency windows. Several runs were performed from different starting points. A rough manual tuning of the parameters was sufficient to ensure convergence of the algorithm to the same point in each run.

A calibrator, shown in Fig. 3, was identical to the through-line of the resonators. The measured transmission of the calibrator and cables shows very linear behavior within each frequency window. Therefore, a linear loss vs frequency model was used to compensate for the cable loss.

## 3 Results

Typical measured and computed insertion loss results are shown in Fig. 4. Only the resonances in the 5.5–5.9 GHz and 7.5–7.9 GHz windows were used in optimizing the model to determine the loss of the resonator. Due the high-Q of the circuit at 9 GHz, the $S_{21}$ data in that window did not have enough points to permit a good fit using the optimizer.

The measured attenuation constant of the resonant transmission line at 300 K and 4.2 K is shown in Fig. 5. The values shown are the equivalent attenuation constant $\alpha_0$ at $f_0 =$ 0.1 GHz, assuming the scaling law $\alpha(f) = \alpha_0 (f/f_0)^{1/2}$. The theoretical value of the attenuation constant at 300 K is also shown for reference; it was estimated from the bulk DC conductivity $\sigma$ and a simple geometrical calculation using the skin depth formula $\delta = 1/(\pi\mu\sigma f)^{1/2}$ to calculate the conductor resistance per unit length.

A commonly used property of a conductor is the residual resistivity ratio, RRR, defined as (DC resistance at 290 K)/(DC resistance at 4.2 K). If the relationship between the electrical conductivity $\sigma$ of the conductor and its microwave loss is known, it is possible to deduce RRR from the microwave loss measurements at room temperature and 4.2 K.

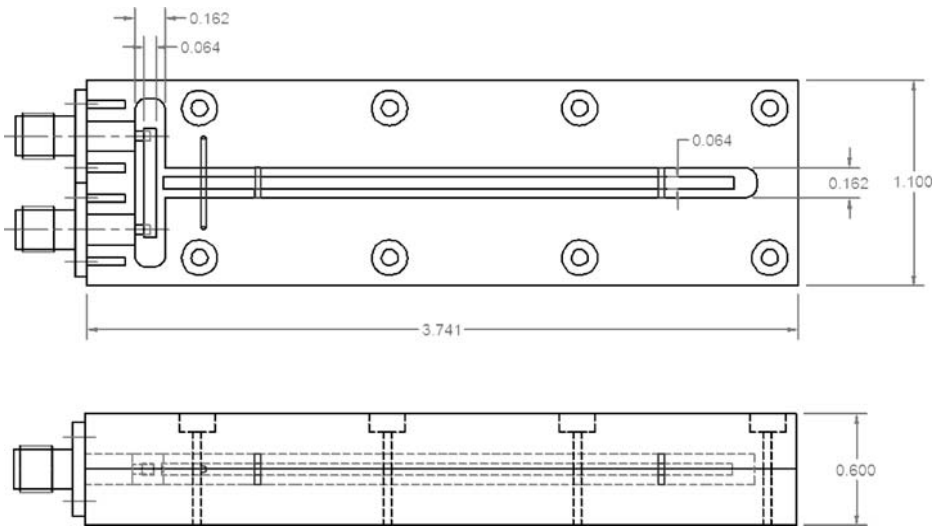

**Fig. 1** The resonator design. The body and center conductor are made of the same material. The center conductor is supported on HDPE pins. Dimensions in inches.

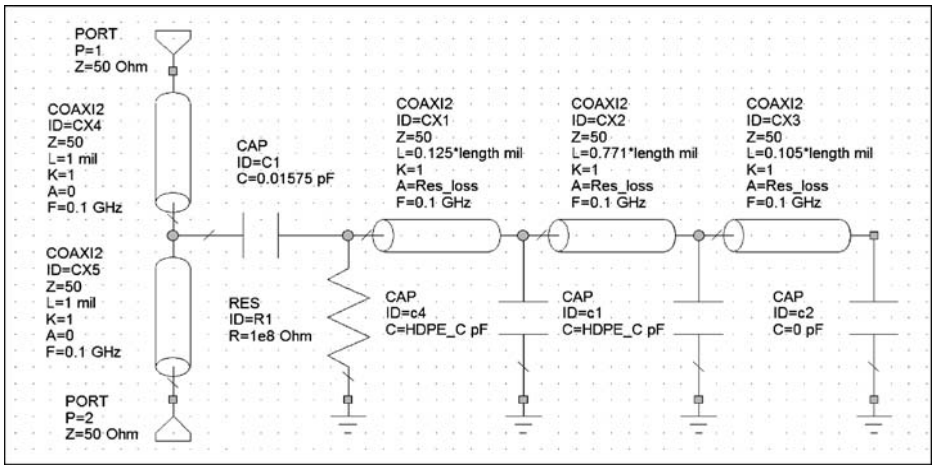

Fig. 2 Equivalent circuit of the resonator.

Figure 6 shows the standard transmission line circuit model. R, L, G, and C are the per-unit-length transmission line parameters, $R_0$ is the real part of the characteristic impedance $Z_0$, and $G_0$ the real part of the characteristic admittance $Y_0=1/Z_0$.

Considering the very low conductivity of the HDPE support pins, $G/G_0$ is taken as zero. According to the classical skin-depth formula $\delta = (1/\pi\sigma\mu f)^{1/2}$, the RF resistance R per unit length of the line is inversely proportional to the conductivity and the skin-depth, so we can write $R=C/\sigma\delta$ or $R=C'/\sigma^{1/2}$. Therefore,

$$\frac{R_{300K}}{R_{4K}} = \sqrt{\frac{\sigma_{4K}}{\sigma_{300K}}} = \frac{\alpha_{300K}}{\alpha_{4K}} \quad \text{or} \quad \left(\frac{\alpha_{300K}}{\alpha_{4K}}\right)^2 = \frac{\sigma_{4K}}{\sigma_{300K}} = RRR$$

RRR calculated in this way assumes the classical skin effect applies. In fact, this may not be the case in materials of high conductivity at low temperature in which the electron mean free path is comparable to or greater than the skin depth, the so-called *anomalous* skin effect regime [5] in which the loss is higher than expected from the classical skin depth. Also, the value of RRR calculated from the measured loss applies only to the conductor material near the surface where machining damage is likely to be greatest, thereby giving a smaller local RRR than in the bulk material. Figure 7 shows the residual resistivity ratio calculated as above for the tested materials.

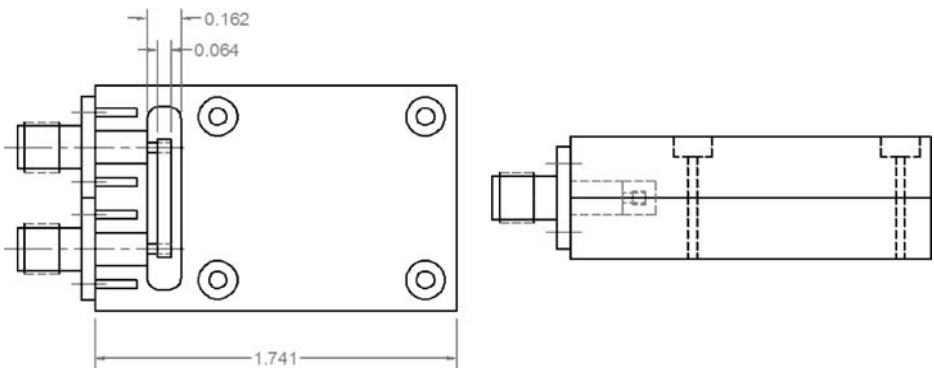

Fig. 3 Calibrator. Dimensions in inches.

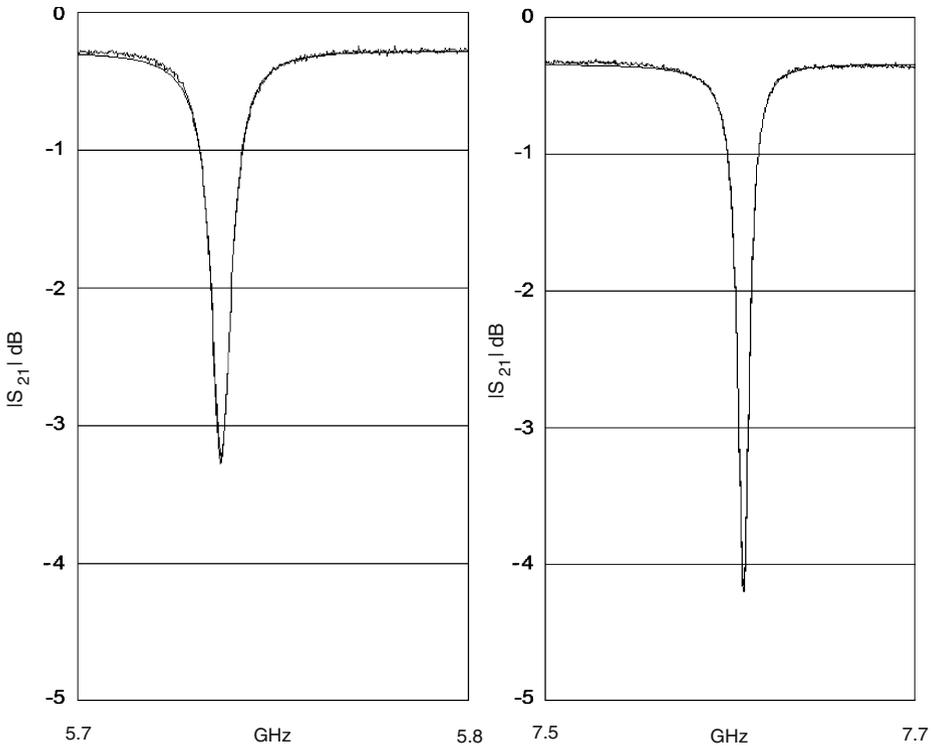

**Fig. 4** Typical measured and simulated |S$_{21}$|. The measured data are normalized to the measured loss of the calibration piece and cables. The simulated data (smooth curves) are almost indistinguishable from the measured data.

## 4 Discussion: the anomalous skin effect

As mentioned in the previous section, when the mean free path of the charge carriers is comparable to or greater than the skin depth, the loss of the conductor is higher than would be calculated using the classical skin depth formula. This phenomenon, the anomalous skin effect [5], limits the degree of RF loss reduction obtained by cooling, particularly in high conductivity metals.

Using the free electron model we can calculate the mean free path as a function of the metal conductivity. Consider the expression for the current density $J$ in terms of the electron drift velocity $V_d$ and the external electric field $E$:

$$\vec{J} = ne\vec{V}_d \quad (1)$$

$$\vec{J} = \sigma\vec{E}, \quad (2)$$

where $n$ denotes the free electron density, $e$ the electron charge and $\sigma$ the electrical conductivity. The drift velocity, which is the result of applying an external electric field, can be expressed in term of the accelerating field $E$ and the mean time between collisions $\tau$ of the electrons as follows:

$$\vec{V}_d = \frac{e\vec{E}}{m}\tau, \qquad (3)$$

where $m$ is the electron mass and $E$ varies slowly compared with $\tau$.

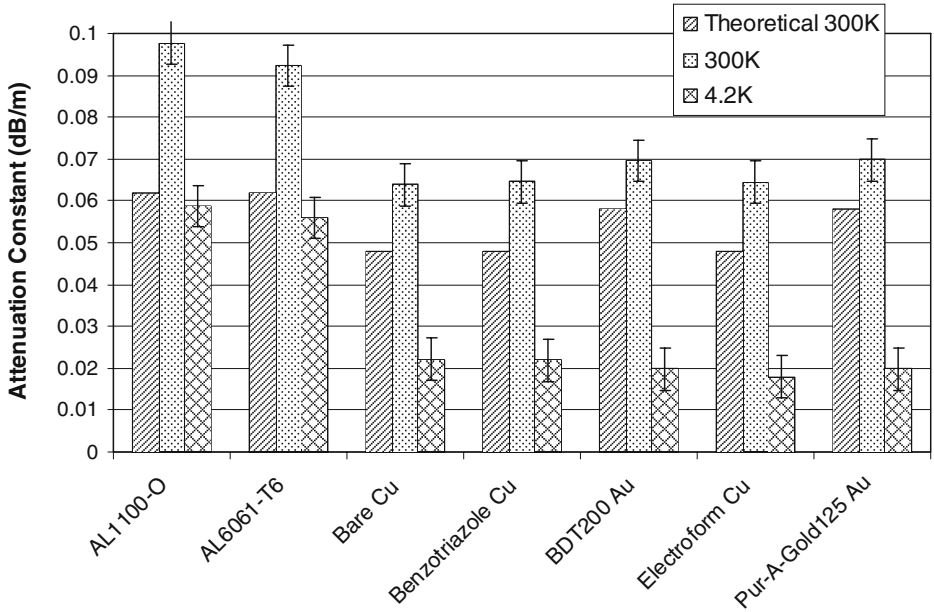

**Fig. 5** Equivalent attenuation constant $\alpha_0$ (dB/m) at 0.1 GHz. The theoretical value at 300 K is shown for reference.

In metals, the conduction electrons belong to the partially filled valence band and therefore can be considered to be at the Fermi level $E_f$ except at high temperatures when it is necessary to take into account the Fermi-Dirac energy distribution for the electron gas.

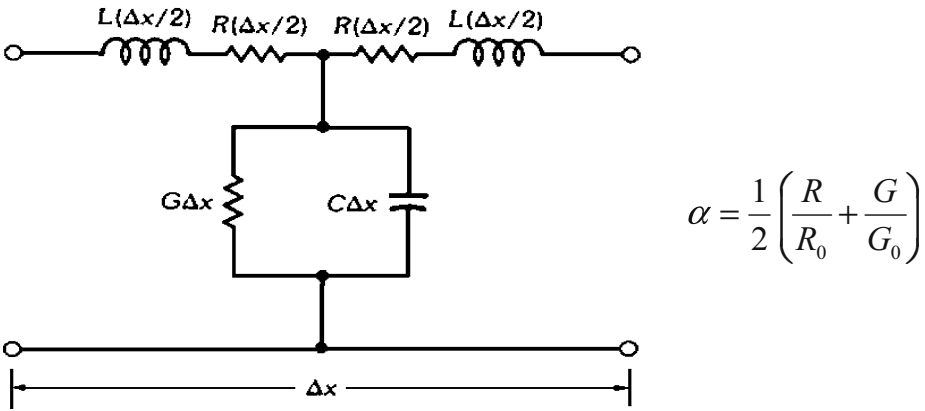

$$\alpha = \frac{1}{2}\left(\frac{R}{R_0} + \frac{G}{G_0}\right)$$

**Fig. 6** Standard transmission line circuit model.

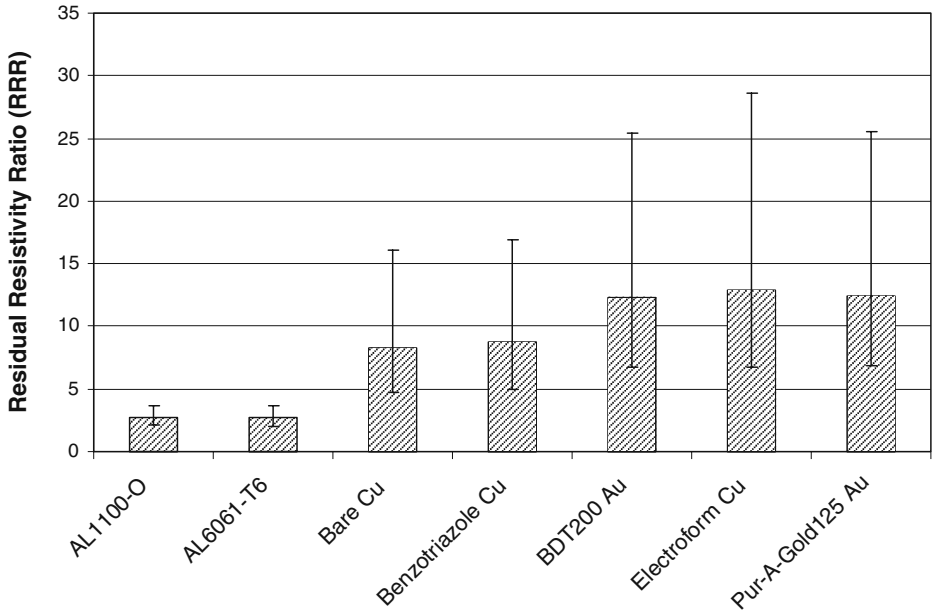

**Fig. 7** Residual resistivity ratio as deduced from the classical skin effect formula. This assumes the conductors are not in the anomalous skin effect regime.

With this consideration the mean time between collisions $\tau$ can be expressed in terms of the Fermi speed $V_f = (2E_f/m)^{1/2}$ and the mean free path $\ell$:

$$\tau = \frac{\ell}{V_f}. \quad (4)$$

Equations (1)–(4) allow the mean free path $\ell$ to be expressed in terms of the Fermi speed $V_f$, the free electron density $n$ and the DC conductivity $\sigma$ as

$$\ell = \frac{\sigma m V_f}{n e^2}. \quad (5)$$

Also, the conductivity $\sigma$ can be expressed in terms of mean free path and Fermi speed:

$$\sigma = \frac{\ell n e^2}{m V_f}. \quad (6)$$

Table 1 shows the Fermi energy, Fermi speed and the free electron density for the most common conductors.

4.1 C101 copper at 4 K

DC resistivity measurements on a machined C101 copper bar gave RRR=106 [7], *i.e.*, the sample has a DC conductivity 106 times grater at 4 K than at 300 K; specifically, $\sigma_{300K} = 5.85 \cdot 10^7$ S/m and $\sigma_{4K} = 6.20 \cdot 10^9$ S/m. With these conductivities, the classical skin depth at

**Table 1** Fermi energy, fermi speed and free electron density of the most common conductors.

| Conductor | Fermi Energy ($E_f$) eV | Fermi Speed ($V_f$) $10^6$ m/s | Free Electron Density (n) $10^{28}/m^3$ |
|---|---|---|---|
| Cu | 7.00 | 1.57 | 8.47 |
| Ag | 5.49 | 1.39 | 5.86 |
| Au | 5.53 | 1.40 | 5.90 |
| Al | 11.70 | 2.03 | 18.10 |

From ashcroft and mermin [6].

6 GHz is $\delta_{300K}=8.50\cdot10^{-7}$ m and $\delta_{4K}=8.25\cdot10^{-8}$ m, and the mean free path calculated using the equation (5) is $\ell_{300K} = 3.9 \cdot 10^{-8}$ m and $\ell_{4k} = 4.1 \cdot 10^{-6}$ m. Therefore:

$$\frac{\ell_{4K}}{\delta_{4K}} = 49.5 \quad \text{and} \quad \frac{\ell_{300K}}{\delta_{300K}} = 0.045.$$

The above calculation shows that, at microwave frequencies and cryogenic temperatures, high purity copper is in the extreme anomalous skin effect regime described by Pippard [5].

4.2 Free electron gas in the extreme anomalous regime

The effective number of electrons that contribute to the conduction process in the extreme anomalous limit ($\ell \gg \delta$) is of order $n_{eff} = (\delta_a/\ell)n$ where $\delta_a$ is the anomalous skin depth. The effective conductivity can also be defined as $\sigma_{eff} = (\delta_a/\ell)\sigma$ and turns out to be independent of the mean free path [8]:

$$\sigma_{eff} = \frac{\delta_a}{\ell}\sigma = \frac{ne^2\delta_a}{mV_f}. \qquad (7)$$

Relating $\delta_a$ and $\sigma_{eff}$ in the same manner as in the non-anomalous regime, $\delta_a = 1/(\pi\mu\sigma_{eff}f)^{1/2}$.

Therefore,

$$\sigma_{eff} = \left(\frac{n^2e^4}{\pi\mu m^2 V_f^2 f}\right)^{\frac{1}{3}} \qquad (8)$$

For C101 copper, the effective conductivity in the anomalous limit at 6 GHz is $\sigma_{eff}= 4.6*10^8$, which yields RRR=7.9, close to the value of 8.3 as measured in the 5–7 GHz band, and far below the DC value of RRR=106 measured for the same material.

## 5 Conclusions

For the copper and gold-plated resonators, the microwave loss decreased by a factor of ~3 on cooling from room temperature to 4 K. This may be sufficient to improve the input loss of a receiver to the point where connector loss dominates. Benzotriazole treatment of copper does not appear to affect the loss significantly.

For the two aluminum resonators, the loss reduction on cooling was less than a factor of 2, resulting in a cold loss about three times greater than that of the copper or gold conductors.

Benzotriazole passivated copper may be suitable for some RF applications such as large waveguide parts.

Aluminum plated with copper or gold could have the advantage of low cold loss with low weight and desirable mechanical properties.

For the C101 copper conductor, the residual resistivity ratio deduced from the microwave loss measurements, assuming classical skin effect, is smaller than the DC resistance ratio. This is primarily a result of the extreme anomalous skin effect, although it may in part be due to machining damage to material within the skin depth [9]. The free electron gas theory of conduction [8] predicts a resistivity ratio in the extreme anomalous limit close to the measured value.

It is not necessary to cool a conductor all the way to 4 K to benefit from reduced microwave loss. Most of the reduction of resistivity occurs well above 4 K; for copper and gold, a factor of 10 improvement has occurred by 50 K [1, 10].